\documentclass[twocolumn,prb,secnumarabic,citesort,showpacs]{revtex4}%
\usepackage{amsfonts}
\usepackage{amsmath}
\usepackage{amssymb}
\usepackage{graphicx}

\newcommand{\Bn}{B_{\mathrm{N}}}
\newcommand{\Bp}{B_{||}}
\newcommand{\Btot}{B_{\mathrm{tot}}}
\newcommand{\cond}{e^{2}/h}
\newcommand{\gavg}{G_{\mathrm{avg}}}

\newcommand{\pI}{I(\mathbf{r},t)}
\newcommand{\Tp}{T_{\mathrm{P}}}
\newcommand{\tR}{\tau_{\mathrm{R}}}
\newcommand{\uV}{\mu \mathrm{V}}
\newcommand{\um}{\mu \mathrm{m}}
\newcommand{\Vdc}{V_{\mathrm{dc}}}

\begin{document}
\title{Nuclear Polarization in Quantum Point Contacts in an In-Plane Magnetic Field}
\author{Y. Ren,$^{1}$ W. Yu,$^{1}$ S. M. Frolov,$^{1, \ast}$ J. A. Folk$^{1}$ and W. Wegscheider$^{2}$}
\affiliation{$^{1}$Department of Physics and Astronomy, University of British Columbia, Vancouver, BC V6T 1Z1, Canada\\
$^{2}$Laboratorium f\"{u}r Festk\"{o}rperphysik, ETH Z\"{u}rich, 8093 Z\"{u}rich, Switzerland}
\begin{abstract}
Nuclear spin polarization is typically generated in GaAs quantum point contacts (QPCs) when an out-of-plane magnetic field gives rise to spin-polarized quantum Hall edge states, and a voltage bias drives transitions between the edge states via electron-nuclear flip-flop scattering. Here, we report a similar effect for QPCs in an in-plane magnetic field, where currents are spin polarized but edge states are not formed. The nuclear polarization gives rise to hysteresis in the d.c.~transport characteristics, with relaxation timescales around 100 seconds. The dependence of anomalous QPC conductance features on nuclear polarization provides a useful test of their spin-sensitivity.

\end{abstract}
\date{\today}


\maketitle

QPCs are the simplest of all semiconductor nanostructures: short one-dimensional constrictions between regions of two-dimensional  electron gas (2DEG), with conductance quantized in units of $G=2\cond$ at zero magnetic field and low temperature (the factor of two comes from spin degeneracy), or $1e^2/h$ at high magnetic field when spin degeneracy is broken.\cite{Wees, Wharam} Despite their apparent simplicity, the spin physics of QPCs has inspired a great deal of debate in the last 10 years, ever since it was pointed out that their low-conductance transport characteristics ($G<2e^2/h$) deviate from a simple non-interacting picture.\cite{Tarucha, Yacoby, Thomas}

One of these anomalous characteristics is a zero-bias conductance peak (ZBP) observed for $G<2\cond$ at low temperature. As the applied magnetic field is increased, some ZBPs collapse without splitting while others split into two peaks by $2g\mu_{\mathrm{B}}B$, with Land\'{e} $g$-factor ranging from much less than the bulk value in GaAs, $g= 0.44$, to much greater than $0.44$.\cite{Cronenwett, Sarkozy, Chen}  The complicated field dependence of ZBPs has given rise to controversial explanations, ranging from Kondo physics\cite{Cronenwett, Rejec} to a non-spin-related phenomenological model.\cite{Chen}

Despite extensive experimental and theoretical work to understand the electron spin physics of QPCs below the $2\cond$ plateau, the effects of
nuclear spin on QPC conductance have only been studied deep in the quantum Hall regime, where many of the conductance anomalies disappear.\cite{Wald, Corcoles} Over the last decade, however, it has become increasingly clear that understanding the electron spin physics of semiconductor nanostructures requires a careful consideration of the influence of nuclear spin via the hyperfine interaction.\cite{Ono, Koppens}  This is especially true for nanostructures defined in GaAs and other III-V materials, where large atomic masses lead to a large electron-nuclear coupling constant through the Fermi contact interaction.\cite{Paget0}

The hyperfine interaction gives rise to an effective magnetic field acting on electron spin that is proportional to the local nuclear spin polarization.  Significant nuclear polarizations can be built up, also via the hyperfine interaction, when a nonequilibrium population of electron spins relaxes, flipping nuclear spins to conserve angular momentum: a process known as dynamic nuclear polarization (DNP).\cite{Abragam}  For example, a large d.c.~bias applied between spin-polarized edge states in the quantum Hall regime often leads to DNP, which can then change transport characteristics significantly. This mechanism has been used to generate and detect nuclear polarization in QPCs, giving rise to hysteresis in the conductance as a function of d.c.~bias.\cite{Wald}

In this paper, we show that DNP in QPCs is not limited to the quantum Hall regime: it is a generic feature of nonequilibrium transport through a QPC in an external magnetic field.  The magnetic fields applied in this work were primarily in the plane of the electron gas; the out-of-plane component was much too small to give rise to Landau quantization. Nuclear polarization is manifested by hysteresis in the differential conductance as a function of d.c.~bias across the QPC, with polarization and relaxation timescales that are consistent with other nanostructures in GaAs.\cite{Wald, Baugh, Reilly_arxiv} DNP leads to changes in conductance features such as ZBPs, which sheds light on the origin of those conductance features by confirming a dependence of the ZBP on electron spin physics in the QPC.\cite{Cooper}

\begin{figure}[t]
  \includegraphics[scale=1.0]{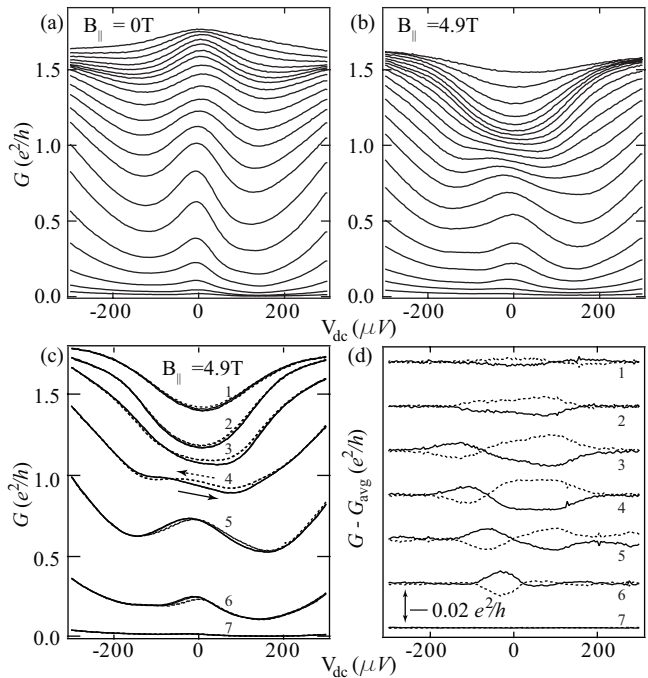}\\
  \caption{QPC nonlinear conductance and hysteresis. (\textbf{a}) Nonlinear conductance $G$ versus d.c.~bias $\Vdc$  for a range of gate voltages, at $\Bp=0$T and  (\textbf{b}) $\Bp=4.9$T.
  (\textbf{c}) Nonlinear conductance curves showing hysteresis for $\Vdc$ swept from  negative to positive (solid) and from positive to negative  (dashed). Labels 1-7 denote select curves from (b).
  (\textbf{d}) Hysteresis measured as deviation from average conductance (described in text) for corresponding curves 1-7 in  (c). Arrow  indicates  vertical scale.}
  \label{}
\end{figure}

Three $1\um$-long and four $0.5\um$-long QPCs were defined using electrostatic gates on the surface of a GaAs/AlGaAs heterostructure. The gates were made using electron beam lithography, and the lithographic width of the QPC (the distance between electrostatic gates) was $0.25\mu$m. The 2DEG was located 110 nm below the surface, with a mobility of $\mu=4.4\times10^{6}\mathrm{cm}^{2}/\mathrm{Vs}$ measured at $T=1.5$K. The experiment was performed at a electron temperature of $40$mK in a dilution refrigerator.  A magnetic field up to $12\mathrm{T}$ was applied within $0.5^\circ$ of the plane of the sample, so the primary effect from the field was through the induced spin-splitting. Differential conductance, $G\equiv dI/dV_{\mathrm{bias}}$, was measured using a lock-in with a d.c.~source-drain bias superimposed on top of a $10\uV$ a.c.~excitation, $V_{\mathrm{bias}}=\Vdc+V_{\mathrm{ac}}$.

Nonlinear conductance characteristics were typical of reports in the literature for low-disorder QPCs, both at zero field (Fig.~1a) and high field
(Fig.~1b).  A clear ZBP was observed throughout the range  $2\cond\gtrsim G \gtrsim  0.01\cond$ at zero field.  The ZBP collapsed at high field, and
a spin-resolved plateau appeared at $1\cond$.

Looking more closely at the high field data, however, the differential conductance was slightly dependent on sweep direction (Fig.~1c), giving rise to hysteresis in traces of $G$ as $\Vdc$ was swept from negative voltage to positive and back to negative. The sweep rate used to gather these hysteresis curves was $8\mu$V/s, chosen to be fast enough that relaxation was minimized during the sweep, but slow enough that artificial hysteresis due to a lag in instrumentation was eliminated. Subtracting off the average of the two sweep directions, $\gavg$, the hysteresis is visible primarily within the range $|\Vdc|\lesssim 200\uV$, corresponding to the bias window where the ZBP is observed (Fig.~1d).

\begin{figure}[t]
  \includegraphics[scale=1.0]{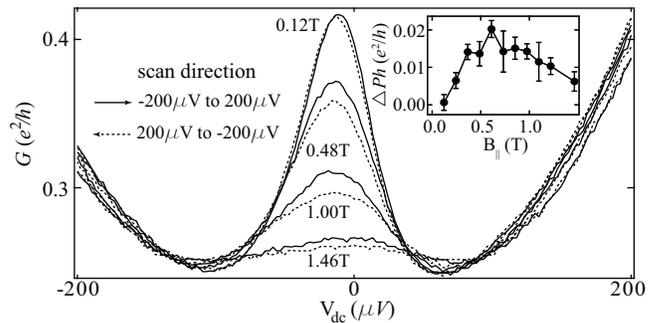}\\
  \caption{Evolution of ZBP and hysteresis in an in-plane magnetic field $\Bp$ from 0.12T to 1.46T (different QPC to that used for Fig.~1).
  Inset: $\Bp$ dependence of peak-height difference $\Delta Ph$ extracted from curves in main panel}
  \label{}
\end{figure}

The magnetic field and time dependence of the hysteresis provide insight into its origin (Figs.~2 and 3). Hysteresis was absent at zero magnetic field, then grew with field on a scale of hundreds of mT. The fact that hysteresis appears only in the presence of a field suggests that it is related to spin instead of, for example, to thermal effects or bias-induced switching of charged dopant sites.

Hysteresis timescales were measured by applying $|\Vdc|\gtrsim 50\mu$V for a time $\Tp$, then switching $\Vdc$ rapidly to zero and monitoring the conductance over several minutes. Fig.~3a shows relaxation traces associated with $\Tp$ from 20s to 1800s. The conductance relaxes over time in all traces, with a typical time constant on the order of 100s. This time constant is much longer than relaxation times associated with electron spin in GaAs, but consistent with previous reports of nuclear spin relaxation in GaAs 2DEG nanostructures.\cite{Wald, Baugh, Reilly_arxiv} These relaxation measurements suggest that the hysteresis cannot be due to electron spin polarization, leaving nuclear spin as a likely explanation.

The observation of similar hysteresis and conductance relaxation effects in all QPCs measured demonstrates that nuclear polarization is a generic feature for QPC nonlinear transport in an in-plane magnetic field.  When the nuclear polarization is small, the change in zero-bias conductance away from its equilibrium value, $G_{eq}$, can be assumed to be proportional to nuclear polarization.  A careful analysis of the conductance change during the nuclear spin relaxation process can therefore shed light on the relaxation mechanisms involved.

\begin{figure}
  \includegraphics[scale=1.0]{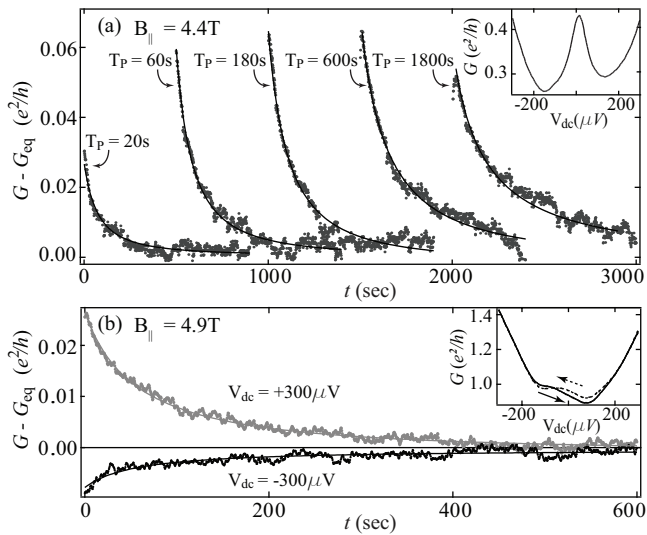}\\
  \caption{(\textbf{a}) Relaxation of zero-bias conductance away from its equilibrium value, $G_{eq}$, for various $\Tp$ at $\Bp=4.4$T (dot). Solid lines are fits of data to Eq.~(\ref{relaxation}). Curves are offset horizontally by 500s for clarity. Inset: zero-bias peak measured at the same gate voltage and $\Bp$.
  (\textbf{b}) Fits of Eq.~(\ref{relaxation}) (solid) to conductance relaxation curves (dot) of another QPC defined in the same wafer. Black and grey curves represent polarizing bias $\Vdc=-300\uV$ and $+300\uV$, respectively. Inset: hysteresis measured at the same gate voltage and $\Bp$.}
  \label{}
\end{figure}

Nuclear spin relaxation in GaAs nanostructures occurs through spin-lattice relaxation, coupling with conduction electrons (Korringa relaxation) and by spin diffusion. Qualitative evidence for the importance of the diffusion mechanism can be found in the dependence of relaxation time on the high-bias polarization time, $\Tp$, as shown in Fig.~3a. The signal measured immediately after returning to zero bias ($t=0$) grows with $\Tp$ up to $\Tp=180$s but saturates and even begins to decrease for longer polarization times. This implies that longer polarization times lead initially to larger nuclear spin polarization but, for $\Tp\gtrsim 180$s, the polarization rate is matched by spin-lattice relaxation and out-diffusion.  However, relaxation curves associated with longer $\Tp$ consistently relax more slowly, indicating that the area over which nuclei are polarized continues to increase with polarization time and providing evidence for nuclear spin diffusion.

Spin diffusion and spin-lattice relaxation can be included in the dynamics of nuclear spin polarization $\pI$ in a 3-dimensional system by solving:
\begin{equation}\label{relaxation}
    \partial \pI/\partial t=D\nabla^2_{\mathbf{r}}\pI-\pI/\tR+S(\mathbf{r},t),
\end{equation}
where the first term on the right-hand side represents nuclear spin diffusion with uniform rate $D$ and the second term represents spin-lattice
relaxation with characteristic time $\tR$.  Korringa relaxation effectively adds to the second term in Eq.~(\ref{relaxation}), but the Korringa rate is expected to be an order of magnitude slower than the spin-lattice rate at 40mK.\cite{Berg}  The last term, $S(\mathbf{r},t)=I_0 \eta(\mathbf{r})[\theta(t)-\theta(t-\Tp)]$, describes the QPC as a spatially localized source of nuclear spin polarization during $0<t<\Tp$, with geometry $\eta(\mathbf{r})$ and polarization rate $I_0$.

Approximating the QPC as a spherical source of nuclear polarization with radius $a$, $\eta(\mathbf{r})=e^{-\mathbf{r}^2/a^2}$, Eq.~(\ref{relaxation}) can be solved analytically.  The parameters $I_0$, $\tR$, and $a$ were fit to the data for 5 different polarization times (Fig.~3a) using published values for the nuclear spin diffusion constant in GaAs, $D\sim 10^{-13}$cm$^2$/s.\cite{Paget}  $a$ and $\tR$ were constrained to be the same for all polarization times because they are related to GaAs material properties and the QPC itself. $I_0$ was fit independently for each curve, reflecting the possibility that  the polarization rate may depend on $\Tp$.\cite{Petta, Reilly_arxiv}  The spin-lattice relaxation time extracted from these fits, $\tR=2600\pm 100$s, is consistent with NMR measurements of spin-lattice relaxation times in GaAs.\cite{McNeil}  It is an order of magnitude longer than decay times seen in Fig.~3a, however, indicating that the contribution of spin-lattice relaxation is overwhelmed by that of spin diffusion.

The value of $a$ extracted from the fits is related to the volume over which polarization occurs.  For the QPC in Fig.~3a, $a$ was found to be $65\pm5$nm, similar to the Fermi wavelength $\lambda_\mathrm{F}=75$nm in the bulk 2DEG. Data from a second QPC defined on the same wafer yielded $a=42\pm5$nm (Fig.~3b), using the same $\tau_R$ and $D$.  In practice, however, Eq.~(\ref{relaxation}) ignores spatial dependence in the spin diffusion rate, which would be expected due to different electron densities.\cite{Deng}  Furthermore, the geometry of the polarizing region, $\eta(\mathbf{r})$, is not known. It must not be thicker than the electron wavefunction transverse to the 2DEG, i.e., a disk rather than a sphere, and it may be closer to a rod than to a disk due to the 1D geometry of the QPC itself.  More complicated $\eta(\mathbf{r})$ require numerical solutions to Eq.~(\ref{relaxation}).  Taking into account numerical solutions of Eq.~(\ref{relaxation}) for a range of reasonable geometries,  the dimensions of polarization source must all be less than $\sim5\lambda_\mathrm{F}$, indicating that polarization happens in the QPC rather than in the leads.

\begin{figure}
  \includegraphics[scale=1.0]{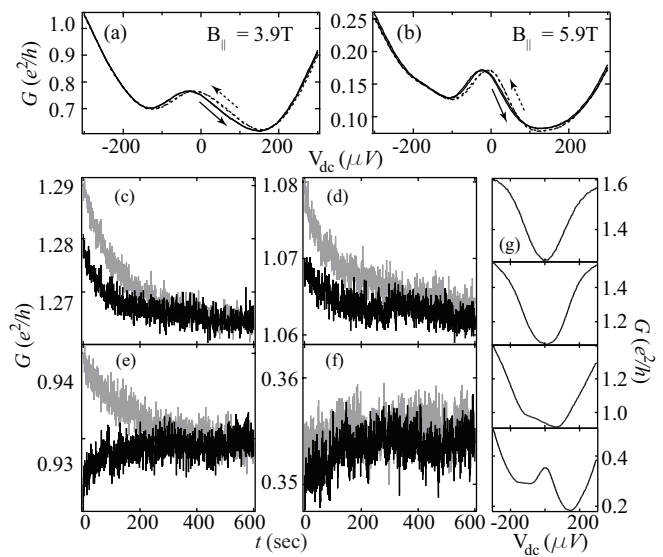}\\
  \caption{
  (\textbf{a}, \textbf{b}) Examples of hysteresis effect that are more complicated than those shown in Fig.~2.
  (\textbf{c})-(\textbf{f}) A transition between symmetric and antisymmetric conductance relaxation curves at $\Bp=4.9$T as readout conductance decreases from $G=1.27e^2/h$ to $0.35e^2/h$ while keeping the conductance during the polarization step unchanged at $1.27e^2/h$. Black and grey curves correspond to $\Vdc=-300\uV$ and $+300\uV$, respectively.
  (\textbf{g}) Zero-bias peaks measured at the same readout gate voltage as in (c)-(f).}
  \label{}
\end{figure}

The effect of nuclear polarization in Fig.~2 is similar to that of the external field: it creates an additional nuclear field, $\Bn=\Btot-\Bp$, in the total effective magnetic field, $\Btot$, as one might expect from a uniformly polarized 2DEG.  The direction of $\Bn$ depended on the polarity of the polarizing bias (see also Fig.~3b), perhaps due to asymmetry in the QPC potential.  Comparing the magnitude of the hysteresis with the effect of $\Bp$, an estimate $\Bn\sim 0.1$T at $\Bp=0.48$T can be extracted from Fig.~2.  Values up to $\Bn\sim 0.3$T were found for other gate settings and other $\Bp$.  Another similarity between the effects of nuclear polarization and in-plane field in this measurement is the insensitivity of the conductance to DNP at high bias.  In the conductance regime explored in this paper (below the $2e^2/h$ plateau) the external field was observed to have a strong effect on the ZBP ($|\Vdc|\lesssim 200\uV$) but only a weak effect at higher bias. The absence of hysteresis outside $|\Vdc|\lesssim 200\uV$ can then be explained by an insensitivity of the differential conductance to spin effects at low conductance and high d.c.~bias.

The hysteresis behavior in Fig.~2, in which DNP mimics external field, is consistent with the assumption of Eq.~(\ref{relaxation}) that nuclear polarization is generated at one site in the QPC, then diffuses to the surrounding area creating a broad region of polarization aligned uniformly parallel or anti-parallel to the external field. More complicated hysteresis curves require more sophisticated explanations, however. In Figs.~4a and 4b, the conductance peak shifts to slightly positive or negative bias depending on the sweep direction, but does not change significantly in height. A shift in the total effective magnetic field due to nuclear polarization cannot, by itself, explain this behavior.  Taking all measured QPCs into account, no consistent correlation was observed between the type of hysteresis (as in Fig.~2 vs Figs.~4a and 4b) and the external parameters such as field, gate voltage, etc.

Conductance relaxation measurements showed a non-trivial bias dependence that is similarly difficult to explain with a single field of nuclear polarization. For a given sign of bias (positive or negative), nuclear polarization depended only slightly on $|\Vdc|$ for $|\Vdc|\gtrsim50\uV$.  Changing the sign of the applied bias, on the other hand, sometimes led to significant changes.  Relaxation curves for QPCs polarized under opposite biases were always symmetric for $1\cond\lesssim G\lesssim 2\cond$.  In Fig.~4c, for example, positive and negative bias polarizations both decrease in conductance as they relax.  At lower conductance, the curves were often antisymmetric: as in Figs.~3b and 4e, positive and negative bias polarizations often relaxed in opposite directions.

The origin of symmetric and antisymmetric behaviors was investigated by separating polarization and readout processes: polarizing at one conductance, rapidly changing the gate voltage to give a different conductance when bias was removed, then measuring relaxation at the new conductance.
A transition from symmetric, to antisymmetric, and back to symmetric relaxation curves can be seen in Fig.~4c-f, covering a range of readout conductances but maintaining the same polarization conductance.  Analogous transitions between symmetric and antisymmetric relaxation curves occurred for all QPCs, and at many different fields, but the readout conductance where the transitions occurred varied widely.  In some QPCs the antisymmetry remained for polarization and/or readout conductance down to $\sim 0.1e^2/h$. These data indicate that the same polarization state can lead to relaxation curves with arbitrary sign depending on the readout conductance, despite the fact that an external $B_{||}$ always decreased the conductance.  This observation is impossible to explain with a single field of nuclear polarization.  It is necessary, therefore, to consider a more sophisticated model of nuclear spin configuration in the QPC and the microscopic mechanism of how it is generated under high bias.

\begin{figure}[t]
  \includegraphics[scale=1.0]{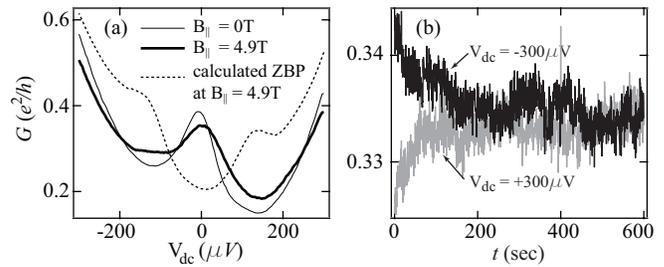}\\
  \caption{
    (\textbf{a}) Comparison between measured ZBP at $\Bp=4.9$T, and the expected curve based on a simple Kondo-like splitting. The dashed line was calculated by taking the ZBP at $\Bp=0$T and splitting it by $2g\mu_{\mathrm{B}}\Bp$, using $g=0.44$.
    (\textbf{b}) Zero-bias conductance relaxation curves measured at settings for panel (a) ($\Bp=4.9$T), indicating dependence of zero-bias-peak height on nuclear polarization in the QPC.}
  \label{}
\end{figure}

Unlike measurements in the quantum Hall regime, there are no edge channels in an in-plane magnetic field, so the flip-flop scattering that leads to DNP cannot be due to scattering between spin-polarized edge channels as described in Ref.~\onlinecite{Wald}. Instead, another flip-flop mechanism known as spin injected dynamic nuclear polarization (SIDNP) may be responsible for the present data.\cite{Aronov, Johnson}  SIDNP was studied in ferromagnetic (FM)/nonmagnetic (NM) heterostructures,\cite{Strand} where spin-polarized current injected from the FM layer creates nonequilibrium spin magnetization in the NM layer and  the nonequilibrium magnetization polarizes nuclei via the hyperfine interaction.  Nonequilibrium magnetization can also be generated by injecting electrons through a spin-polarized QPC, then transferred to nuclei in the vicinity of the QPC by DNP.

Considering both source and drain contacts, injection through a spin-polarized QPC more closely resembles a NM/FM/NM system than the NM/FM system from Ref.~\onlinecite{Strand}---essentially two SIDNP junctions back to back.  Applying a voltage to drive electrons from source to drain across a spin-up polarized QPC creates an excess of spin-up electrons in the drain, while leaving an excess of spin-down electrons in the source.  Flip-flop relaxation of spin-up and spin-down electrons creates opposite nuclear polarizations.  One might therefore expect opposite nuclear polarizations on either end of the QPC, leading to a dipole field acting on conduction electrons. A full explanation for bias-dependent nuclear polarization would have to take into account this dipole field, as well as device-dependent asymmetries in the QPC itself.

The in-plane field dependence of QPC conductance features has historically been used to discern whether or not they originate from spin-related effects. When the in-plane field dependence is simple (e.g. features split with a voltage corresponding to the Zeeman energy), this is a useful tool.  But large in-plane fields also affect orbital electron characteristics, and the connection to spin is ambiguous when experimental data cannot be easily correlated with Zeeman energy.

Nuclear polarization, on the other hand, affects only the spin degree of freedom for electrons moving through a QPC. For this reason, the question of whether a particular conductance feature is spin-related can be answered by observing whether it is affected by DNP.  As an example, the splitting of the ZBP in Fig.~5 does not follow the predictions of a Kondo model in a simple way.\cite{Cronenwett}  The field dependence is very weak up to $\sim 5T$, and the dominant feature in the 4.9T data is a ZBP that is {\em not} split by magnetic field.  One might, therefore, attribute the ZBP entirely to non-spin-related phenomena.  But from the dependence of the ZBP height on nuclear polarization (Fig.~5b), a degree of spin-dependence in the feature can be confirmed. This observation does not rule out an additional contribution to the ZBP that does not depend on spin.  A recent theoretical proposal in Ref.~\onlinecite{Cooper} suggests that more detailed measurements of the nuclear relaxation time may further narrow the range of possible explanations for conductance anomalies in QPCs.

This work was supported by NSERC, CFI, and CIFAR. W.W. acknowledges financial support by the Deutsche Forschungsgemeinschaft (DFG) in the framework of the program ¡®¡®Halbleiter-Spintronik¡¯¡¯ (SPP 1285).

\end{document}